\documentclass[aps,prl,twocolumn,superscriptaddress,longbibliography]{revtex4-1}

\usepackage{graphicx}
\usepackage{amsmath}
\usepackage{xcolor}

 \newcommand{\bra}[1]{\left< #1\right|}
\newcommand{\ket}[1]{\left|#1 \right>}

\begin{document}
 
\title{Signature of preformed pairs in angle-resolved photoemission spectroscopy}

\author{Klemen Kova\v c}
\thanks{These authors contributed equally to this work.}
\affiliation{J. Stefan Institute, 1000 Ljubljana, Slovenia}
\affiliation{Faculty of Mathematics and Physics, University of Ljubljana, 1000 Ljubljana, Slovenia}

\author{Alberto Nocera}
\thanks{These authors contributed equally to this work.}
\affiliation{\!Department \!of \!Physics and Astronomy, \!University of\!  British Columbia, \!Vancouver, British \!Columbia,\! Canada,\! V6T \!1Z1}
\affiliation{\!Stewart Blusson Quantum Matter \!Institute, \!University  of British Columbia, \!Vancouver, British \!Columbia, \!Canada, \!V6T \!1Z4}

\author{Andrea Damascelli}
\affiliation{\!Department \!of \!Physics and Astronomy, \!University of\!  British Columbia, \!Vancouver, British \!Columbia,\! Canada,\! V6T \!1Z1}
\affiliation{\!Stewart Blusson Quantum Matter \!Institute, \!University  of British Columbia, \!Vancouver, British \!Columbia, \!Canada, \!V6T \!1Z4} 

\author{Janez Bon\v ca}
\affiliation{J. Stefan Institute, 1000 Ljubljana, Slovenia}
\affiliation{Faculty of Mathematics and Physics, University of Ljubljana, 1000 Ljubljana, Slovenia}

\author{Mona Berciu}
\affiliation{\!Department \!of \!Physics and Astronomy, \!University of\!  British Columbia, \!Vancouver, British \!Columbia,\! Canada,\! V6T \!1Z1}
\affiliation{\!Stewart Blusson Quantum Matter \!Institute, \!University  of British Columbia, \!Vancouver, British \!Columbia, \!Canada, \!V6T \!1Z4}

\begin{abstract}
We use density matrix renormalization group (DMRG) and variational exact diagonalization (VED) to calculate the single-electron removal spectral weight for the Hubbard-Holstein model at low electron densities. Tuning the strength of the electron-phonon coupling and of the Hubbard repulsion allows us to contrast the results for a liquid of polarons versus a liquid of bipolarons. The former shows spectral weight up to the Fermi energy, as expected for  a  metal. The latter has a gap  in its spectral weight, set by the bipolaron binding energy, although this is also a (strongly correlated) metal. This difference suggests that angle-resolved photoemission spectroscopy could be used to identify liquids of pre-formed pairs. Furthermore, we show that the one-dimensional liquid of incoherent bipolarons is well approximated by a 'Bose sea' of bosons that are hard-core in momentum space, 
occupying the momenta inside the Fermi sea but otherwise non-interacting. This new proposal for a strongly-correlated many-body wavefunction opens the way for studying various other properties of  incoherent (non-superconducting) liquids of pre-formed pairs in any dimension.
\end{abstract}

\maketitle
    {\em Introduction:} Strongly correlated electron states are challenging to characterize  both theoretically and experimentally. One very interesting class of such states are liquids of pre-formed pairs, expected to appear in systems where exchange of some type of boson results in binding of electron pairs, however the interplay of their density, the temperature and other factors is such that the pairs are not coherent (no superconductivity). Such a state has been hypothesized to exist between the superconducting temperature $T_C$ and the pseudogap temperature $T^*$ in underdoped cuprates \cite{cuprates1,c2,c3,c4,c5,c51, c6, c61, c62}, although this is debated \cite{Marcel, M1,John}. The undoped ground-states of BaBiO$_3$ and  of rare-earth nickelates were also argued to be crystals of bipolarons \cite{Steve,st2,st21,st3, st4,st41, st42, st43}, {\em i.e.} of pairs of holes bound through electron-phonon coupling, residing on the compressed BiO$_6$ and NiO$_6$ octahedra, respectively. Upon doping, the bipolaron crystal melts into a liquid of bipolarons. Recently, bipolaron formation due to phonon zero-point fluctuations was found in a model with a quadratic electron-phonon coupling~\cite{Kivelson2024}. Such binding can also be enhanced by the resonant driving of the system. Extensive theoretical work has been conducted on this topic~\cite{Subedi2014, Babadi2017, John2021, Kovac2024} to explain numerous experiments where intriguing nonthermal phases of matter were observed following resonant driving with laser pulses~\cite{Rini2007, Hu2014, Buzzi2020}.

    Direct experimental detection of  such preformed pairs remains a challenge, as is the formulation of a theoretical description of such a state. The extraordinary success of angle-resolved photoemission spectroscopy (ARPES) for characterizing single quasiparticle properties, has inspired a considerable effort to find generalizations that might provide direct spectroscopic signatures of pair formation. So far, the focus has been on detecting a pair of ejected electrons, whether from one \cite{2e1,2e1a,2e1b,2e1c} or two \cite{2e2,2e2a} incident photons. While these ideas are promising, such higher order processes are more difficult to detect.

    In this Letter, we show that  ARPES already contains clear signatures of preformed pair formation. We prove this by studying a one-dimensional (1D) Hubbard-Holstein model using  density matrix renormalization group (DMRG)\cite{Schollwock2011}. Changing the ratio of the electron-phonon coupling {\em vs.} the electron-electron Hubbard repulsion allows us to vary the ground-state from a liquid of unbound polarons (quasiparticles whereby the carrier is dressed by a cloud of phonons) to a liquid of bipolarons (bound pairs of polarons). The DMRG calculated spectral weights prove that the latter is signalled by the appearance of a 'gap' (equal to half the binding energy) below the Fermi energy, even though this phase is metallic. We explain this behavior using a simple theoretical picture of the bipolaron liquid as a 'Bose sea' of hard-core bosons {\em in momentum space}, which is validated by comparison against the variationally exact DMRG results.

    We emphasize that we expect this ARPES signature of pre-formed pairs to be generic. The Holstein coupling is convenient because of the multiple tools already developed to study it~\cite{Wellein1997, Jeckelmann1998,  Bonca1999, Barisic2002, Hohenadler2003, Berciu2006, Mitric2022, Mitric2023}, but we expect the polaron and bipolarons phenomenology to remain similar for other types of bosons and electron-boson couplings in all dimensions. Similarly, we do not expect dimensionality to play a role in the proposed ansatz for the liquids of incoherent pre-formed pairs, except that in higher-D these would occur at finite temperatures. Here we consider a 1D chain for access to easier-to-converge DMRG calculations and because its ground-state cannot be ordered (superconducting) \cite{Hoh,Mermin-Wagner}.

\begin{figure*}
  \includegraphics[width=2\columnwidth]{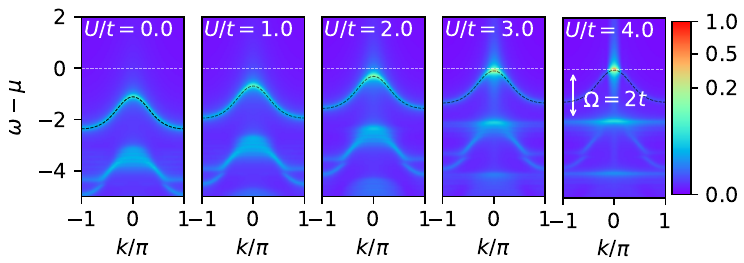}
  \caption{\label{fig1} Spectral weight $A(k,\omega)$ for a system with $N_e=2$ electrons on $N=29$ sites, at fixed $t=1$, $\Omega=2, \lambda=1$ and increasing $U=0,\dots, 4$ (left to right panels) computed with VED. The white horizontal dashed line marks the Fermi energy $\omega = \mu=E_P(0)-\Delta/2$ (see text for further discussion) while the black dashed line shows the inverted polaron dispersion, $-E_P(k)$ (up to an overall shift). As $U$ increases, the $N_e=2$ GS evolves from a bipolaron (spectra display a  'gap' between the lowest binding energy feature and $\mu$) to two unbound polarons (spectral weight starting at $\mu$). See text for more details. }
\end{figure*}

{\em Model:} We study the 1D Hubbard-Holstein model:
\begin{align}
  \nonumber
{\cal H} = &-t \sum_{i, \sigma}\left(c^{\dagger}_{i\sigma} c_{i+1,\sigma}+ h.c.\right) + U\sum_i {\hat n}_{i\uparrow} {\hat n}_{i\downarrow} \\
  \label{m1} & + \Omega \sum_i b^\dagger_i b_i + g \sum_i  {\hat n}_{i} \left(b_i + b_i^\dagger \right)
\end{align}
where $c^{\dagger}_{i\sigma}$ and  $b_i^\dagger$ create an electron with spin $\sigma$ and a boson, respectively, at site $i=1,\dots,N$, where $N$ is the number of sites and we set $a=1$, $\hbar=1$.  The number operator for electrons with spin-$\sigma$ is ${\hat n}_{i\sigma}=c^{\dagger}_{i\sigma}c_{i\sigma}$, and ${\hat n}_{i}=\sum_\sigma{\hat n}_{i\sigma}$. The first  term describes the bare  electron hopping,  the second is the on-site Hubbard repulsion,  the third is the energy of the Einstein phonons, and the last is the Holstein coupling, which we characterize in terms of the dimensionless electron-phonon coupling parameter $\lambda=g^2/(2t\Omega)$, equal to the ratio between the lattice deformation energy $g^2/\Omega$ in the atomic limit, and half
the non-interacting bandwidth $2t$.

We study this model (i) for systems with $N_e=1,2$ electrons on a finite chain with periodic boundary conditions employing variational exact diagonalization (VED)~\cite{Bonca1999, Bonca2000, Ku2002} (we verified that the results are converged and correspond to the thermodynamic limit $N\rightarrow \infty$), and (ii) for finite electron concentrations $N_e/N=n_{\rm dop}< 1$ in finite-size chains with open boundaries, using DMRG directly in the frequency space, with a recently developed variant~\cite{Nocera2022} of the correction-vector approach \cite{Kuhner1999,Jeckelmann2002,NoceraPRE2016}. 
As customary \cite{AndreaRMP}, the ARPES intensity is assumed to be proportional to the spectral weight:
\begin{align}
  \label{m2}
  A(k,\omega) =- {1\over \pi} \text{Im}\left[\sum_{\alpha,\sigma} \frac{|\langle\alpha, N_e-1|c_{k\sigma}|GS, N_e\rangle|^2}{\omega +i\eta + E_{\alpha}^{(N_e-1)}-E_{GS}^{(N_e)}} \right]
\end{align}
Here, ${\cal H }|\alpha, N_e\rangle = E_{\alpha}^{(N_e)} |\alpha, N_e\rangle$ are the eigenstates with $N_e$ electrons, labelled by their quantum numbers $\alpha$. The lowest-energy eigenstate is the ground-state (GS). 
Technical details about the evaluation of $A(k,\omega)$ using VED and DMRG are in the Supplementary Material \cite{SM}.
 
{\em Results:} We first consider an infinite chain with $N_e=2$  electrons in its GS, to see the change in $A(k,\omega)$ when the two carriers are bound (at large $\lambda$, small $U$) {\em vs.} unbound (at small $\lambda$, large $U$). Either way,  upon photo-ejection of an electron with momentum $k$, the system is left with one electron and total  momentum $-k$. The spectrum $E_{\alpha}^{(1)}$ is well known, see Fig.~\ref{figA1} of End Matter: the lowest eigenstate
is the polaron band with energy $E_P(k)$, located below the polaron+one-phonon continuum starting at $E_P(0)+\Omega$, and other, even higher energy features~\cite{Bonca1999, Berciu06, Bonca2019}. 

Using Eq. (\ref{m2}) and the fact that $E_{\alpha}^{(1)}\ge E_P(k)$, we expect the feature in $A(k,\omega)$ with the lowest binding energy to appear at energy $\omega(k) = E_{GS}^{(2)}-E_P(k)$,  and thus it must disperse like the inverted polaron band   $-E_P(k)$, regardless of whether the GS has two unbound polarons and $E_{GS}^{(2)}= 2E_P(0)$, or a bipolaron and $E_{GS}^{(2)}= 2E_P(0)-\Delta$, with $\Delta$ being the binding energy.

Furthermore, in this canonical calculation the chemical potential is $\mu= [E_{GS}^{(N_e+1)}-E_{GS}^{(N_e-1)}]/2$. If the GS has unbound polarons, then $\mu = E_P(0)$ and we find $\omega(k) \le \omega(0)=\mu$, {\em i.e.} no gap is expected in the spectral weight. On the other hand, if the GS has a bipolaron then $\mu= E_P(0)-\Delta/2$ and we find  $\omega(k) \le \omega(0)=\mu-\Delta/2$, {\em i.e.} a gap equal to $\Delta/2$ is expected in the spectral weight.

These expectations are confirmed in Fig. \ref{fig1}, where we plot $A(k,\omega)$ at $\lambda=1, \Omega=2t$ for $U/t=0,\dots,4$. We shift energies 
so that  the  'Fermi energy'  is at the usual ARPES $\omega-\mu$=0 value, shown by the horizontal line.

Electron-removal weight is indeed seen only below $\omega-\mu$=0. (The vertical 'flashlight' visible above is due to the finite $\eta$ in $A(k,\omega)$). In all cases, the dispersion of the lowest binding energy feature indeed follows  $-E_P(k)$ (up to a vertical shift). The latter is shown by the  black dashed line, and is extracted from the polaron band in Fig.~\ref{figA1}. For $U\leq2$, this lowest binding energy feature lies at a finite energy below the $\omega - \mu$=0  line, as if the system is gapped (``insulating''), and this gap indeed equals $\Delta/2$ as expected for a bipolaron GS. For $U\gtrsim 3$ the Hubbard repulsion is too strong and the ground-state is a ``metal''  of two unbound polarons (we use quotation marks because the electron  density vanishes in the thermodynamic limit). The ARPES spectral weight starts at $\omega-\mu=0$, as expected for a 'metallic' GS.

\begin{figure}
  \includegraphics[width=\columnwidth]{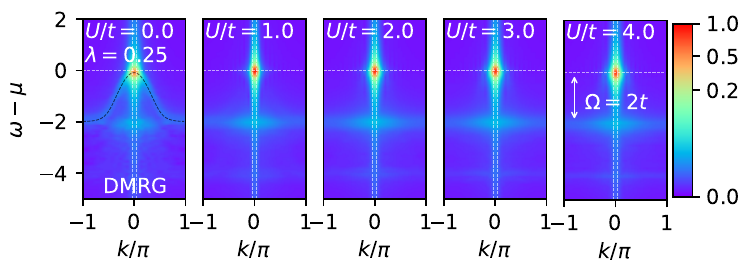}
    \includegraphics[width=\columnwidth]{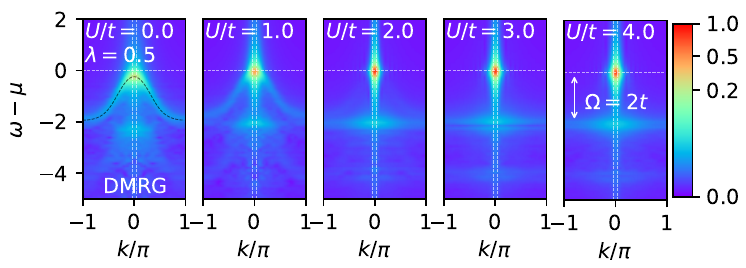}
    \includegraphics[width=\columnwidth]{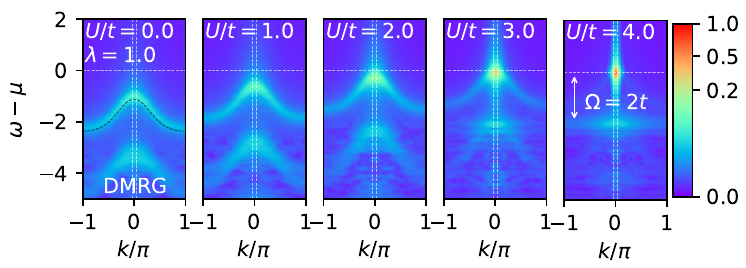}
    \includegraphics[width=\columnwidth]{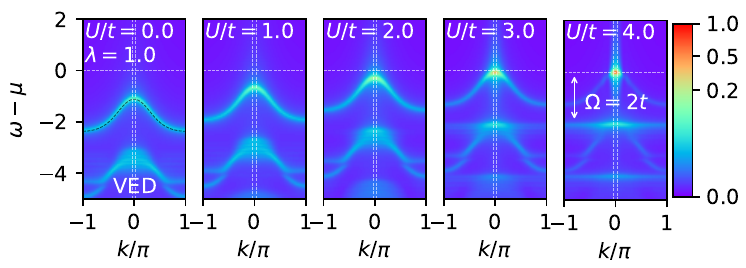}
      \caption{\label{fig2} The upper three rows show $A(k,\omega)$ calculated with DMRG for $n_{\rm dop}=0.0833$ and $t=1,\Omega=2$, $\eta=0.15$. {DMRG simulations were performed on a $N=48$ sites chain with open boundary conditions and $N_e=4$.} The onsite repulsion $U=0\rightarrow 4$ increases from left to right, while the effective el-ph coupling $\lambda=0.25, 0.5, 1$ increases from top to bottom. The lowest row shows $A(k,\omega)$ assembled from single pair VED results like those of Fig. 1, as explained in the text. {Each single pair VED calculation was conducted on a system with $N = 29$.  For the VED simulations, $\eta$ was set to 0.05, while all other parameters remained consistent with those used in the DMRG calculations.} In all panels, the dashed horizontal line is the Fermi energy $\omega=\mu$, while vertical dashed lines indicate the Fermi momenta $\pm k_F$. }
\end{figure}

Figure \ref{fig1} confirms that in $A(k,\omega)$, the key signature indicating the existence of a pre-formed pair (the bipolaron) is the appearance of a gap below the Fermi energy; this gap closes for unbound quasiparticles (polarons).
{\em This is the first main result of this work}. Before continuing, we comment on a secondary but important difference between the two types of GS. The left (bipolaron) panels show spectral weight at all $k$ in the lowest-binding energy feature that traces  (the shifted) $-E_P(k)$, whereas in the right (unbound polarons) panels, the spectral weight in this feature is strongly concentrated at $k=0$. This is because the bipolaron GS wavefunction $|BP\rangle \sim [ \sum_k \phi_k c^\dagger_{k\uparrow} c^\dagger_{-k\downarrow} + \sum_{k,q} \phi_{k,q} c^\dagger_{k\uparrow} c^\dagger_{-k-q,\downarrow} b^\dagger_q+\dots]|0\rangle$ describes binding in real space. As a result, $c_{k\sigma}|BP\rangle\sim [\phi_k  c^\dagger_{-k,-\sigma} +\sum_q \phi_{k,q} c^\dagger_{-k-q,-\sigma} b^\dagger_q+\dots]|0\rangle$ has a finite overlap at all $k$ values with the single polaron eigenstate with momentum $-k$ and spin $-\sigma$, defined as $|P, -k, -\sigma\rangle = [ \sqrt{Z_k} c^\dagger_{-k,-\sigma} +\sum_q \psi_{k,q} c^\dagger_{-k-q,-\sigma} b^\dagger_q+\dots]|0\rangle$  ($Z_k$ is the quasiparticle weight). In contrast, the GS with unbound polarons  is described well as an uncorrelated pair of GS polarons with opposite spins and zero momentum $|P, 0, \uparrow\rangle\otimes|P, 0, \downarrow\rangle$. The ARPES weight now is like that from of a single polaron \cite{Bonca2019, Bonca2022} with a strong peak at $\omega-\mu=0, k=0$, plus spectral intensity  at $\omega -\mu=  - n \Omega$ for all $k$, indicating the $n\ge 1 $ phonons left behind after the electron was photoejected. Indeed, we verified that the  ARPES weight for the $N_e=1$ polaron GS is identical to that in the right panel of Fig. \ref{fig1}, proving that the second polaron is just a bystander.

Next, we demonstrate that similar ARPES spectra are obtained for low, finite electron concentrations $n_{\rm dop}$.
The top three rows of Figure \ref{fig2} show ARPES spectra calculated with DMRG at $n_{\rm dop}=0.0833$, for $\lambda=0.25, 0.5, 1$ (top to bottom, respectively). Columns correspond to $U/t=0\rightarrow 4$ (left $\rightarrow$ right, respectively).  In all cases the spectral weight of the lowest binding-energy  feature again follows $-E_P(k)$, see dashed black lines in the first column. A gap equal to $\Delta/2$ is observed in the panels with small $U$ and large $\lambda$, where the GS is a liquid of bipolarons. {\em We emphasize that this gap does  not imply an insulating ground state}: the bipolaron liquid is a metal, albeit a strongly  correlated one. The gap means that single carrier fluctuations are gapped, however two-carrier (pair) fluctuations are not. We verify this by showing that the Drude conductivity is finite for both types of liquids, see Appendix B of End Matter.

The gap in the single-electron removal spectral weight closes as $U$ increases and/or $\lambda$ decreases and the bipolarons unbind, forming a liquid of polarons with momenta $|k|\le k_F$. Indeed, here the ARPES weight for the lowest binding energy feature vanishes for $|k|> k_F$, as expected for a normal metal.  Higher binding energy features again reveal the multi-phonon sidebands at $\omega-\mu=-n\Omega$.

\begin{figure}[t]
\includegraphics[width=1\columnwidth]{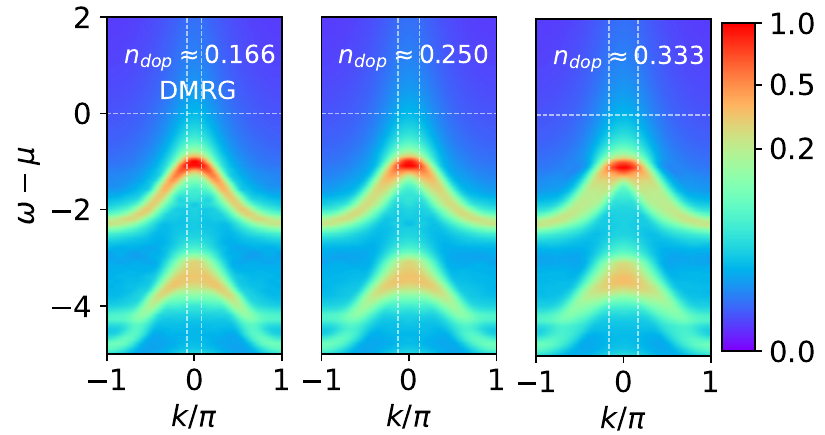}

\includegraphics[width=1\columnwidth]{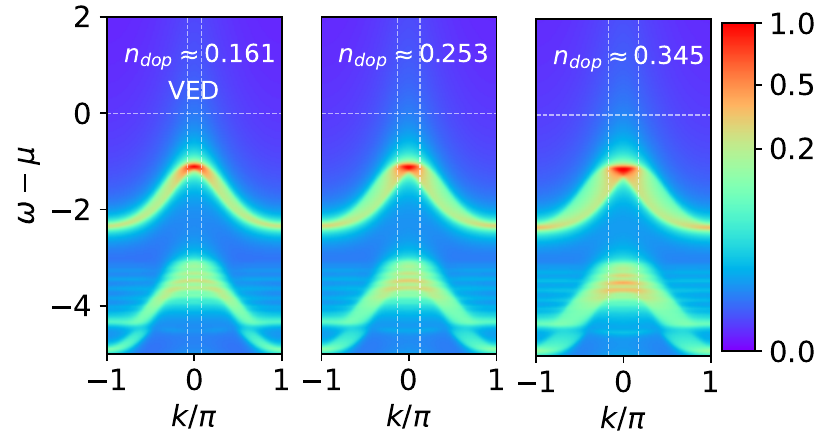}

      \caption{\label{fig3} Comparison between ARPES spectra calculated with DMRG for $n_{\rm dop}=0.166,0.250, 0.333$ (first row, left to right) with those generated from single-pair VED calculations for a liquid of hard-core bosons at the closest attainable doping levels (second row). DMRG simulations were performed on a $N=48$ sites chain with open boundary conditions, while VED calculations were performed on a $N=29$ sites chain with periodic boundary conditions. Other parameters are $U=0, t=1, \Omega=2,\lambda =1, \eta=0.15$ (DMRG), $\eta=0.05$ (VED). See text for more details.}
\end{figure}

The third row panels  look similar to those shown for the same parameters but  $n_{\rm dop}\rightarrow0$ in Fig. \ref{fig1}. The key difference is that when $n_{\rm dop}>0$, the lowest binding energy feature is flat for all $|k| \le k_F= \pi n_{\rm dop}/2$ (marked by vertical dashed lines). This is even more apparent in Fig. \ref{fig3}, which shows results for larger $n_{\rm dop}$. This demonstrates that these bipolarons do not all occupy the bipolaron GS with total momentum $K=0$, consistent with the fact that a superconducting GS is forbidden in 1D systems. \cite{Mermin-Wagner}
If macroscopic occupation of a single level (boson condensation) is forbidden, then the GS must consist of pairs occupying the low-lying bipolaron states with a microscopic occupation number.  We find that we can mimic $A(k,\omega)$ for a {\em finite concentration} using the ARPES spectra calculated like in Fig. \ref{fig1} for a {\em single} pair of electrons, if we assume that the bipolarons are hard-core bosons in momentum space, {\em i.e.} each momentum $|K| \le k_F$ is occupied by one bipolaron. 
ARPES ejection of an electron with momentum $k$ can then be considered as the sum of ARPES spectral weight for ejecting an electron with momentum $k$ from a single bipolaron, summed over the possible bipolaron momenta $|K|\le k_F$. Calculations for such 'Bose seas' are shown in the lowest row of Fig. \ref{fig2}, and are in excellent quantitative agreement with the DMRG results for the same parameters, shown in the row just above it.
 
This agreement persists up to much higher electron concentrations, as demonstrated in Fig. \ref{fig3} by comparing the DMRG spectra obtained for $n_{\rm dop}=0.166, 0.250$ and $0.333$ with spectra generated from sums of single pair VED results, as described above. The two sets of spectral weights are very similar. Both show the flattening of the lowest binding energy feature for $|k|\le k_F$, as well as an overall broadening of this entire feature with increasing $n_{\rm dop}$. The VED results explain that this broadening is a consequence
of the fact that removing an electron with momentum $k$ from a bipolaron with momentum $K$ leaves behind an electron with momentum $K-k$, whose energy (and therefore binding energy) varies when summing up overall bipolaron contributions, $|K|\le k_F$. 

This demonstration that {\em the liquid of bipolarons is well approximated by  
a 'Bose sea' of hard-core (in momentum-space) bosons with  momenta $|K|\le k_F$ but otherwise very weakly interacting}, is the  second major result of our work. It provides a simple physical picture for understanding this correlated unusual 1D metal of pre-formed pairs, and allows us to speculate, for instance, on how this
single-particle gap closes at a temperature comparable with the bipolaron binding energy. Above this temperature all pairs are unbound, hence we expect a typical metallic ARPES spectral weight starting at the Fermi energy (no gap). As the temperature is lowered and an increasingly larger fraction of polarons bind into pairs, their contribution to the spectral weight becomes gapped. As a result, the total spectral weight should be consistent with the opening of a pseudogap for  $k_B T < \Delta$. 

Support for the speculations in the paragraph above is provided in Fig.~\ref{figC1 in the End Matter}, although more work is needed to validate it. We note that pseudogaps are indeed observed in higher-D phases hypothesized to be liquids of pre-formed pairs, {\em e.g.} in underdoped cuprates. We emphasize that we do {\em not} claim that phonons are responsible for cuprate superconductivity; the low carrier limit of our model does not support antiferromagnetic order, and its 2D bipolarons have $s$, not $d$-wave symmetry. Instead, our claim is that a liquid of preformed pairs would likely exhibit a pseudogap in its ARPES weight at finite $T$,  regardless of the specific type of  boson responsible for binding the carriers.

Experimentally, this could be tested (i) with time resolved ARPES (trARPES) in materials with (pseudo)gaps attributed to pre-formed pairs \cite{AD2}. If pumping breaks some pairs, the time-scale over which the (pseudo)gap reopens plus the direct measurement of $\Delta$, might suffice to both  verify the existence of pre-formed pairs, and to identify the boson(s) involved in the pairing; (ii) Applying hydrostatic pressure and/or uniaxial stress can decrease the $U/t$ ratio \cite{nn}. In a material with sufficiently large $\lambda$ this could induce a crossover from a normal metal to a liquid of pre-formed pairs that could be observed in ARPES; (iii) 
trARPES could be combined with experiments that endeavor to change the state of a system through optical pumping of specific phonon modes. Appearance of superconductivity well above the equilibrium $T_C$ has been reported in a variety of such systems \cite{pp,pp1,pp2,pp3,pp4}. The opening of a (pseudo)gap in trARPES spectra would verify that preformed pairs are  forming,  an important first step towards confirming superconductivity.  Finally, (iv) the evolution of ARPES spectra could most directly be verified in quantum simulators of Hubbard-Holstein models \cite{HHqs,HHqsa}, where parameters can be individually tuned. 

The above suggestions assume that the opening of a (pseudo)gap would be visible in trARPES data, which (i)  might be challenging, and (ii) even if observed, would not rule out other possible explanations. 

The 'Bose sea' of  pairs ansatz we identified in 1D proposes a new type of microscopic wavefunction for liquids of uncondensed pre-formed pairs in any dimension, as a possible intermediary phase between a superconductor with a macroscopic concentration of condensed pairs, and a normal phase where all pairs are unbound. Our work suggests that if the binding energy $\Delta$ is considerable, thermal fluctuations may first excite pairs from the condensate into finite $K$ pair states. The coherence is  lost at $T_C$ where the pairs spread over the Bose sea, leading to an unusual metal of pre-formed pairs (disordered-phase  superconductor).   The breaking of the pairs into single quasiparticles  would occur at much higher temperatures $k_B T \sim \Delta$, leading to a normal metal.

Our 'Bose sea' ansatz for  this unusual metal of pre-formed pairs opens the way to calculate, in future work, its response to  other probes, potentially providing other experimental fingerprints of this type of correlated state. 

{\em Conclusion:} We proposed a new ansatz to describe correlated metals of pre-formed pairs, with 
significant consequences for progress both in the theoretical
understanding of such correlated states, and also
for providing new experimental ways, besides trARPES, to identify the existence of pre-formed pairs and perhaps even of the bosons
responsible for mediating the attraction.

{\em Acknowledgements:} J.B. and K.K. acknowledge the support by the program No. P1-0044 of the Slovenian Research Agency (ARIS). J.B.  acknowledges  discussions with S.A. Trugman, A. Saxena and support from  the Center for Integrated Nanotechnologies, a U.S. Department of Energy, Office of Basic Energy Sciences user facility and Physics of Condensed Matter and Complex Systems Group (T-4) at Los Alamos National Laboratory. This project was undertaken thanks in part to funding from the Max Planck-UBC-UTokyo Center for
Quantum Materials and the Canada First Research Excellence Fund, Quantum Materials and Future Technologies Program, as well as the  Natural Sciences and Engineering Research Council of Canada (A.N., A.D. and M. B.). A.N. acknowledges computational resources and services provided by  Advanced Research Computing at the University of British Columbia.

\newpage\newpage
\section*{End Matter}
\section*{Appendix A: Single polaron spectrum.} 

To determine the polaronic spectrum referenced in the main text, we calculate the spectral function for electron addition to the vacuum state, defined as:

\begin{equation} \label{spektralka}
    A^+(\omega,k) = \sum_{j = 0}^{M}\Big{|}\bra{\alpha, 1}c_{k\sigma}^\dagger\ket{0}\Big{|}^2\\
    \times\delta\Big{(}\omega-E_{\alpha}^{(1)}\Big{)},
\end{equation}
where $\ket{0}$ denotes the vacuum state with zero energy. 

We computed $M = 200$ one-electron states $\ket{\alpha, 1}$, ensuring orthogonality via the Gram-Schmidt reorthogonalization procedure. For a graphical representation of $A^+(\omega,k)$, we used a Lorentzian form of the $\delta$-functions with a half-width at half-maximum of $\eta = 0.05$. 

\begin{figure}[b]
  \includegraphics[width=1\columnwidth]{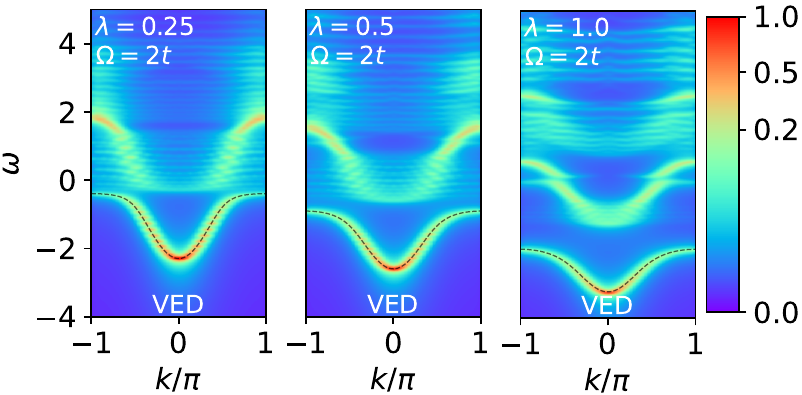}
  \caption{Single polaron spectral weight $A^+(k,\omega)$ for $t=1$, $\Omega=2t$, and increasing $\lambda=0.25,\ 0.5,\ 1.0$ (left to right panels). The black dashed lines trace the corresponding polaron dispersions $E_P(k)$.}\label{figA1}
\end{figure}

Figure~\ref{figA1} shows a well-pronounced quasiparticle peak with energy $E_P(k)$ (dashed line), followed by the incoherent part of the spectrum which begins at $E_P(k) + \Omega$. As expected, larger values of $\lambda$ result in a smaller polaron bandwidth, indicating a heavier Holstein polaron.

\section*{Appendix B: The Drude weight of bipolarons} 

We calculate the Drude weight $\tilde D$ that characterizes the singular contribution to the real part of the conductivity ${\rm Re} \sigma(\omega) = \pi \tilde D \delta(\omega) + \sigma^{\rm reg}(\omega)$, as described in Ref. \cite{Jeckelmann2009}.
Specifically, we use DMRG and VED to evaluate Eqs. (7) and (17) of  Ref. \cite{Jeckelmann2009}, respectively, to obtain $D={\tilde D}/n_{\rm dop}$, {\em i.e.} the contribution per carrier to $\tilde D$.

\begin{figure}[b]
\includegraphics[width=0.95\columnwidth]{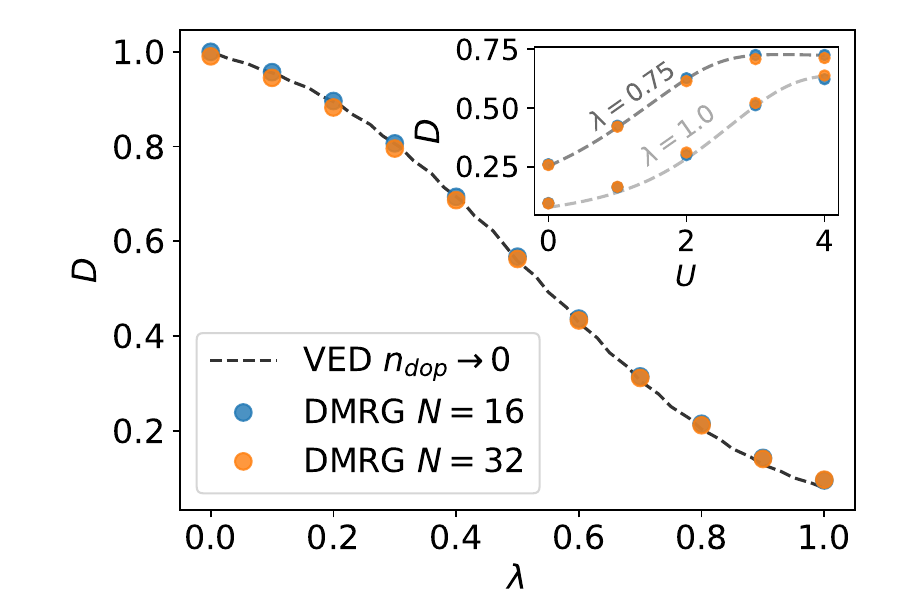}
  \caption{Drude weight per charge carrier $D$ versus $\lambda$ when $U=0$ and the GS is always a bipolaron liquid. The results are obtained with VED (dashed lines) and DMRG (symbols). The latter are for two system sizes and carrier concentration $n_{\rm dop}=0.125$. The inset shows the evolution of $D$ with $U$ for two values of $\lambda$.}\label{figB1}
\end{figure}

We find excellent agreement for the values of $D$ obtained with both methods, for different system sizes and carrier concentrations. A representative subset of our results is presented in Fig. \ref{figB1}. The main panel shows the evolution of $D$ with $\lambda$ when $U=0$, {\em ie} for ground-states that are bipolaron liquids. As expected, $D$ decreases with increasing $\lambda$, reflecting the increased bipolaron mass, but in all cases $D>0$. This verifies that all these bipolaron liquids are metals, consistent with the fact that the bipolarons are charged and mobile. 

The inset of Fig. \ref{figB1} shows the evolution of $D$ with increasing  $U$, for two values of $\lambda$. In both cases, $D$ increases as the bipolarons become less strongly bound, and saturates to the $D$  value of the single polarons for large enough $U$ that the GS is a liquid of unbound polarons.

\section*{Appendix C: Mimicking the pseudogap.} 

The DMRG method implemented here is appropriate for calculating ARPES spectral weights at zero temperature. Nevertheless, we can use it to calculate the spectral weight of a system with a mix of both polarons and bipolarons (thus, mimicking a finite-$T$ system where some bipolarons have been thermally dissociated) by considering a finite chain with an odd number of electrons. For small $U$ and large $\lambda$ most of these electrons pair into bipolarons, but one will be left as a single polaron. If our arguments are correct, the spectral weight for this case should look like a superposition of the spectra of a bipolaron liquid and of a polaron liquid. 

\begin{figure}[b]
  \includegraphics[width=0.52\columnwidth]{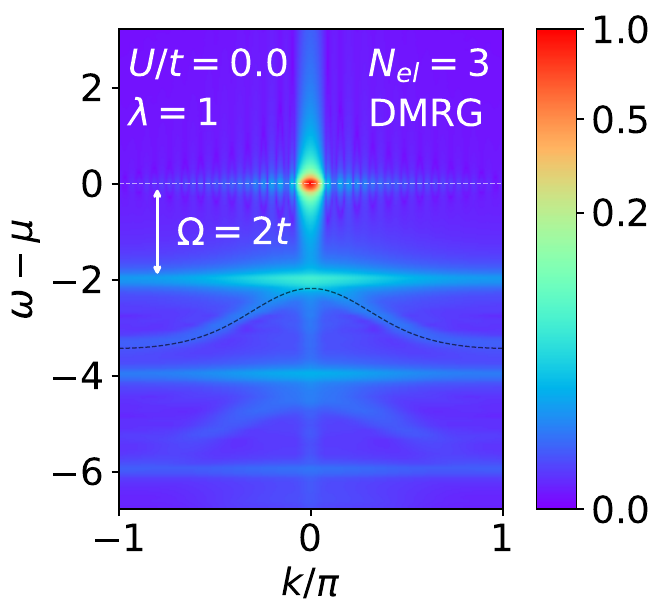}
  \includegraphics[width=0.42\columnwidth]{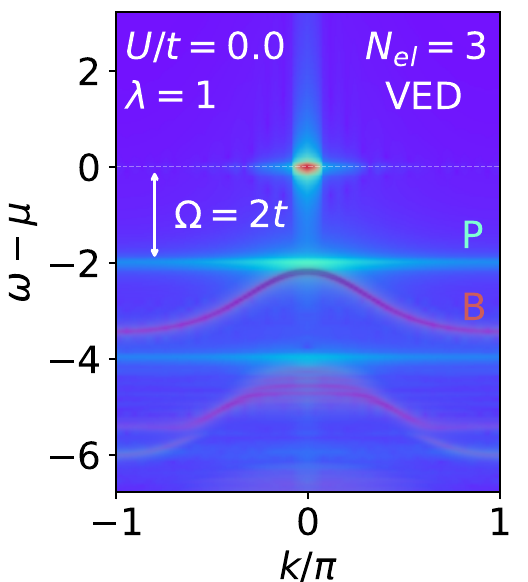}
  \caption{Left: ARPES spectral weight calculated with DMRG for a system with $N_e=3$ electrons, showing the superposition of both single polaron features (ungapped) and bipolaron features (gapped). Right: overlay of the ARPES spectral weights for a single polaron (P, blue colormap) and a single bipolaron (B, pink colormap) calculated with VED for the same parameters. }\label{figC1}
\end{figure}

This is confirmed in Fig.~\ref{figC1}. Its left panel shows the DMRG spectral weight for a chain with $N=48$ sites and $N_e=3$ electrons. As expected, we observe both the spectral weight at $\omega=\mu, k=0$ together with  the flat features at $\omega -\mu= -n\Omega$ expected when the photoelectron is ejected from the single polaron; and also the inverted polaron dispersion $-E_P(k)$ located at a 'gap' $\Delta$ below the Fermi energy, as well as the other features with higher binding energy characteristic of the bipolaron spectral weight.  Note that the gap to the bipolaron signal is different here because we performed a 'spin-polarized' simulation, adding and removing a carrier with the same spin as that of the original unpaired polaron. As a result, a bound pair cannot form in the addition channel, resulting in a 'spin-polarized' $\mu=E_P(0)$. 

This combination of the two types of ARPES signal is made very clear by comparison with the right panel, where we overlayed the ARPES spectral weight for a single polaron (blue colormap) over that of a  single bipolaron (pink colormap). Both were calculated with VED.

\clearpage
\onecolumngrid
\begin{center}
    \textbf{\large Supplemental Material: Signature of Preformed Pairs in Angle-Resolved Photoemission Spectroscopy}\\[0.5em]
    Klemen Kova\v{c}, Alberto Nocera, Andrea Damascelli, Janez Bon\v{c}a, Mona Berciu
\end{center}
\vspace{1em}

\twocolumngrid
\section{Evaluation of the spectral function $A(k,\omega)$ using VED.} 

We employed a numerical method described in detail in Refs. \cite{Bonča1999, Ku2002}. For completeness, we briefly review here the $N_e=2$ solution. A variational subspace is defined on an infinite one-dimensional lattice, and is constructed iteratively beginning with an initial zero-phonon state where both electrons are on the same site.   The variational Hilbert space is then generated from this state by applying $N_h$ times  the sum of the two off--diagonal operators:$\sum_{i, \sigma}\left(c^{\dagger}_{i\sigma} c_{i+1,\sigma}+ h.c.\right) + \sum_i  {\hat n}_{i} \left(b_i + b_i^\dagger \right)$,  taking into account the full translational symmetry. The obtained subspace is restricted in the sense that it allows only a finite maximal distance of a phonon quanta from the doubly occupied site, $L_\mathrm{max_1}=(N_\mathrm{h}-1)/2$, a maximal distance between two electrons $L_\mathrm{max_2}=N_\mathrm{h}$, a maximal amount of phonon quanta at the doubly occupied site $N_\mathrm{phmax}=N_\mathrm{h}$, while at a site that is $L$ sites away from the doubly occupied one,  $N_\mathrm{phmax}=N_\mathrm{h} - 2L$. After the iteration procedure, we exclude the \textit{unused} portion of the chain where electrons cannot reside and where there are no phonon excitations. This results in a finite system of $N=2N_h + 1$ sites, to which we apply periodic boundary conditions. By doing so, we determine the allowed values of $k$ vectors, ensuring that the spectral function is properly defined. We have used a standard Lanczos procedure \cite{Lanczos1950} to obtain the energy spectrum of the model.

The main objective of VED is to analyze the spectral function of the two-electron system corresponding to the removal of an electron with spin $\sigma$. We reformulate the spectral function $A(\omega,k)$ (Eq.~2 of the Main text) in a form more appropriate for numerical calculation as

\begin{multline} \label{spektralka}
    A(\omega,k) = \sum_{j = 0}^{M}\Big{|}\bra{\alpha, 1}c_{k\sigma}\ket{GS, 2}\Big{|}^2\\
    \times\delta\Big{(}\omega+E_{\alpha}^{(1)}- E_{GS}^{(2)}\Big{)},
\end{multline}

We computed $M = 2000$ one-electron states $\ket{\alpha, 1}$, ensuring orthogonality via the Gram-Schmidt reorthogonalization procedure. We used a Lorentzian form of $\delta$-functions with a half-width at half-maximum of $\eta = 0.05$ for graphic representations of $A(\omega,k)$.\\
To improve the precision of density plots, we extended our calculations beyond conventional periodic boundary conditions to incorporate twisted boundary conditions, as discussed in works such as~\cite{Shastry1990, Bonča2003}. Twisted boundary conditions represent a magnetic flux penetrating the ring structure, enabling a continuous connection between discrete $k$ points. Under these conditions, the kinetic energy term in Eq.~1 of the Main text takes the following form

\begin{equation}
    \mathcal{H}_{kin} = -t\sum_{i, \sigma}\left(c^{\dagger}_{i\sigma} c_{i+1,\sigma}e^{i\theta} + h.c.\right).
\end{equation}

For a system of free electrons, this transformation yields the dispersion relation $\varepsilon(k,\theta)=-2t_{el} \cos(k+\theta)$, where $\theta$ represents a magnetic flux that permeates the ring, characterized by $\phi_m =
\theta L/2\pi$ in units of $h/e_0$. Despite the calculations being performed on a finite ring of size $L$, we obtain continuous dispersion relation $\varepsilon(k,\theta)$ by smoothly connecting discrete $k$ points $k_n = 2\pi n/L$ within the interval $\theta \in [0, 2\pi/L]$. 

To validate our spectral function computation, we assessed  (i) its convergence with system size  in order to demonstrate that our method is sufficient to determine the spectral function  in the thermodynamic limit, and (ii) its consistency with theoretical expectations based on the sum rule analysis. We evaluated the integral of $A(\omega, k)$ over the entire frequency range

\begin{equation}    \label{SumRule}
    \int_{-\infty}^{\infty} A(\omega, k) \, d\omega = \bra{GS, 2} c_{k\sigma}^\dagger c_{k\sigma} \ket{GS, 2} = n_{k\sigma}
\end{equation}

Here, $n_{k\sigma}$ represents the electron occupation number with the wavevector $k$ and spin $\sigma$, assuming a complete basis of one-electron states is used. Due to our expanded $k$-point sampling (induced by introducing magnetic flux), we scaled the number of $k$ vectors by a factor $\mathcal{S}$. To obtain the correct sum rule over the entire Brillouin zone: $\sum_{k}n_{k\sigma} = 1$, we have to scale $A(\omega, k) \longrightarrow A(\omega, k) / \mathcal{S}$. 
From here on we set $\mathcal{S} = 3$.

\section{Evaluation of the spectral weight  $A(k,\omega)$ using DMRG.} 

In this section we specify details about the calculations of spectral weight functions with DMRG directly in the frequency space.
For a one-dimensional lattice Hamiltonian of size $N$, it is convenient to reformulate the spectral weight function $A^{\sigma}(k,\omega)$ 
for electron removal with spin=$\sigma$ as
\begin{align}
&A^{\sigma}(k,\omega) = \lim_{\eta\rightarrow 0}\frac{1}{N}\sum_{i,j} e^{i q (i-j)}\times\nonumber\\
&\times -\frac{1}{\pi}\textrm{Im}\Bigg[\langle GS, N_e| \hat{c}^{\dagger}_{i\sigma} \frac{1}{E^{N_e}_{GS}-\hat{H}-\omega+i\eta}\hat{c}_{j\sigma}|GS, N_e\rangle\Bigg],
\end{align}
where one writes down the Hamiltonian propagator explicitly, and $\eta>0$ is an arbitrary small extrinsic spectral broadening.
The correction-vector (CV) method\cite{Kuhner1999,Jeckelmann2002}, 
aims to evaluate, for a specific frequency $\omega$, the real and imaginary part of the correction-vector
\begin{equation}
|x_j(\omega+i\eta)\rangle=\frac{1}{E^{(N_e)}_{GS}-\omega-\hat{H}+i\eta}\hat{c}_{j\sigma}|GS, N_e\rangle
\end{equation}
at fixed finite broadening $\eta$. In this work, we use a method introduced in Ref.~\cite{Nocera2022} 
to compute a generalized correction-vector with smaller entanglement content, the root-$M$ correction vector, defined as
\begin{equation}\label{eq:rootNn1}
	|x^{1/M}_j(\omega+i\eta)\rangle=\Bigg(\frac{1}{E^{(N_e)}_{GS}\omega-\hat{H}+i\eta}\Bigg)^{1/M}\hat{c}_{\sigma j}|GS, N_e\rangle.
\end{equation}
The idea is to construct the actual correction-vector as the final vector
of the series $\{|x^{p/M}_j(\omega+i\eta)\rangle\}_{p\in[1,M]}$
after $M$ applications of the root-$M$ propagator.
In this work, the real and imaginary part of the root-$M$ correction-vector are obtained using a Krylov decomposition of the
Hamiltonian propagator~\cite{NoceraPRE2016}.
The spectral weight function is then computed as a stardard overlap 
$\langle GS, N_e| \hat{c}^{\dagger}_{i\sigma}|x_j(\omega+i\eta)\rangle$ and finally performing a Fourier Transform in momentum space.
For this last step, we use the standard center approximation
\begin{equation}
	A^{\sigma}(k,\omega) \simeq \sum_{i} \cos(k(i-\bar{j}))
	\langle GS, N_e| \hat{c}^{\dagger}_{i\sigma}|x_{\bar{j}}(\omega+i\eta)\rangle,
\end{equation}
where we have indicated with $\bar{j}=N/2$ the center site of the chain.
This approximation reduces the computational cost by order of $N$ and becomes exact in
the thermodynamic limit, although it introduces ``ringing'' artifacts in the spatial Fourier Transform
in small finite systems. As seen in the DMRG results presented in the main paper, 
for systems with  $N=48$ lattice sites the artifacts from the use of open boundary conditions and the center site approximation are minor.

To compute the spectral weight functions, we use the DMRG++ software~\cite{Alvarez0209}, 
using the standard Fock basis representation of the phonon degrees of freedom, utilizing up to $8$ phonon
states to represent the local phonon Hilbert space. We therefore did not use the more sophisticated 
local phonon optimization methods\cite{Jeckelmann1998,DMRG2,Zhang1998,Cheng2012,Brockt2015,Jansen2021,Stolpp2021,DMRGE14,Jansen2022} 
or the recently developed projected purification DMRG method.\cite{kohler2021,mardazad2021}
Agreement between the DMRG and the VED results demonstrates that this approach worked well. 
Excellent comparison between the Momentum Average approximation and the root-$M$ CV method was recently demonstrated in~\cite{Nocera2023}.
For each frequency $\omega$, numerical DMRG simulations were converged with respect to the bond
dimension $m$. A maximum $m = 1024$ (and a minimum $m_{\text{min}}=16$)
provides convergence with a truncation error
smaller than $10^{-7}$ for the frequency dependent calculations.
For the root-$M$ Correction Vector Krylov calculations the choice $M=50$ has shown the
best compromise in terms of moderately large bond dimension required and computational speed.
Finally, we set the Krylov space tridiagonalization error to $\epsilon_{\text{Tridiag}}=10^{-7}$ in order to avoid the proliferation of
Krylov vectors (and thus Lanczos iterations), and their reorthogonalizations.


\begin{thebibliography}{0}%
\makeatletter
\providecommand \@ifxundefined [1]{%
 \@ifx{#1\undefined}
}%
\providecommand \@ifnum [1]{%
 \ifnum #1\expandafter \@firstoftwo
 \else \expandafter \@secondoftwo
 \fi
}%
\providecommand \@ifx [1]{%
 \ifx #1\expandafter \@firstoftwo
 \else \expandafter \@secondoftwo
 \fi
}%
\providecommand \natexlab [1]{#1}%
\providecommand \enquote  [1]{``#1''}%
\providecommand \bibnamefont  [1]{#1}%
\providecommand \bibfnamefont [1]{#1}%
\providecommand \citenamefont [1]{#1}%
\providecommand \href@noop [0]{\@secondoftwo}%
\providecommand \href [0]{\begingroup \@sanitize@url \@href}%
\providecommand \@href[1]{\@@startlink{#1}\@@href}%
\providecommand \@@href[1]{\endgroup#1\@@endlink}%
\providecommand \@sanitize@url [0]{\catcode `\\12\catcode `\$12\catcode `\&12\catcode `\#12\catcode `\^12\catcode `\_12\catcode `\%12\relax}%
\providecommand \@@startlink[1]{}%
\providecommand \@@endlink[0]{}%
\providecommand \url  [0]{\begingroup\@sanitize@url \@url }%
\providecommand \@url [1]{\endgroup\@href {#1}{\urlprefix }}%
\providecommand \urlprefix  [0]{URL }%
\providecommand \Eprint [0]{\href }%
\providecommand \doibase [0]{http://dx.doi.org/}%
\providecommand \selectlanguage [0]{\@gobble}%
\providecommand \bibinfo  [0]{\@secondoftwo}%
\providecommand \bibfield  [0]{\@secondoftwo}%
\providecommand \translation [1]{[#1]}%
\providecommand \BibitemOpen [0]{}%
\providecommand \bibitemStop [0]{}%
\providecommand \bibitemNoStop [0]{.\EOS\space}%
\providecommand \EOS [0]{\spacefactor3000\relax}%
\providecommand \BibitemShut  [1]{\csname bibitem#1\endcsname}%
\let\auto@bib@innerbib\@empty
\end{thebibliography}%


\begin{thebibliography}{99}

\bibitem{cuprates1} V.J. Emery and S.A. Kivelson, \newblock \emph{Importance of phase fluctuations in superconductors with small superfluid density}, \newblock Nature {\bf 374}, 434 (1995), \newblock \doi{10.1038/374434a0}.


\bibitem{c2}D. Nozieres and  S. Schmitt-Rink,\newblock \emph{Bose Condensation in an Attractive Fermion Gas:
From Weak to Strong Coupling Superconductivity}, \newblock J. Low Temp. Phys. {\bf 59}, 195 (1985), \newblock \doi{10.1007/BF00683774}.

\bibitem{c3} Y.J. Uemura, G.M. Luke, B.J. Sternlieb, J.H. Brewer, J.F. Carolan, W.N. Hardy, R. Kadono, J.R. Kempton,
R.F. Kiefl, S.R. Kreitzman, P. Mulhern, T.M. Riseman, D.Ll. Williams, B.X. Yang, S. Uchida, H. Takagi, J. Gopalakrishnan, A.W. Sleight, M.A. Subramanian, C.L. Chien, M.Z. Cieplak, Gang Xiao, V.Y. Lee, B.W. Statt,
C.E. Stronach, W.J. Kossler, and X.H. Yu, \newblock \emph{Universal Correlations between T$_c$ and $n_s/m^*$
(Carrier Density over Effective Mass) in High-T$_c$ Cuprate Superconductors}, \newblock Phys. Rev. Lett. {\bf 62}, 2317  (1989), \newblock \doi{10.1103/PhysRevLett.62.2317}.

\bibitem{c4}Y. J. Uemura, L. P. Le, G. M. Luke, B. J. Sternlieb, W. D. Wu, J. H. Brewer, T. M. Riseman, C. L. Seaman, M. B. Maple, M. Ishikawa, D. G. Hinks, J. D. Jorgensen, G. Saito, and H. Yamochi, \newblock \emph{Basic similarities among cuprate, bismuthate, organic, Chevrel-phase, and heavy-fermion superconductors shown by penetration-depth measurements}, \newblock Phys. Rev. Lett. {\bf 66}, 2665 (1991), \newblock \doi{10.1103/PhysRevLett.66.2665}.

\bibitem{c5}C.A.R. Sa de Melo, M. Randeria, J.R. Engelbrecht, \newblock \emph{Crossover from BCS to Bose superconductivity: Transition temperature and time-dependent Ginzburg-Landau theory}, \newblock Phys. Rev. Lett. {\bf 71}, 3202 (1993), \newblock \doi{10.1103/PhysRevLett.71.3202}.

\bibitem{c51} V.B. Geshkenbein, L.B. Ioffe, and A.I. Larkin, \newblock \emph{Superconductivity in a system with preformed pairs}, \newblock Phys. Rev. B {\bf 55}, 3173 (1997), \newblock \doi{10.1103/PhysRevB.55.3173}.

\bibitem{c6}Shina Tan and K. Levin, \newblock \emph{Nernst effect and anomalous transport in cuprates: A preformed-pair alternative to the vortex scenario}, \newblock Phys. Rev. {\bf B 69}, 064510 (2004), \newblock \doi{10.1103/PhysRevB.69.064510}.

\bibitem{c61} Q. Chen, J. Stajic, S. Tan, and K. Levin, \newblock \emph{BCS-BEC crossover: From high temperature superconductors to ultracold superfluids}, \newblock Physics Reports {\bf 412}, 1 (2005), \newblock \doi{10.1016/j.physrep.2005.02.005}.

\bibitem{c62} Q. Chen, Z. Wang, R. Boyack, and K. Levin, \newblock \emph{Test for BCS-BEC crossover in the cuprate superconductors}, \newblock  npj Quantum Mater. {\bf 9}, 27 (2024), \newblock \doi{ doi.org/10.1038/s41535-024-00640-8}.

\bibitem{Marcel} M. Franz, \newblock \emph{Importance of fluctuations}, \newblock Nature Phys {\bf 3}, 686 (2007), \newblock \doi{10.1038/nphys739}.

\bibitem{M1} I. Hetel, T. Lemberger, T. and M. Randeria, \newblock \emph{ Quantum critical behaviour in the superfluid density of strongly underdoped ultrathin copper oxide films}, \newblock Nature Phys {\bf 3}, 700 (2007), \newblock \doi{10.1038/nphys707}

\bibitem{John} J. Sous,  Y. He, and S. A.  Kivelson, \newblock \emph{Absence of a BCS-BEC crossover in the cuprate superconductors}, \newblock npj Quantum Mater. {\bf 8}, 25 (2023), \newblock \doi{10.1038/s41535-023-00550}. 
  
\bibitem{Steve} B. Lau and A.J.  Millis, \newblock \emph{Theory of the Magnetic and Metal-Insulator Transitions in RNiO$_3$ Bulk and Layered Structures}, \newblock Phys. Rev. Lett. {\bf 110}, 126404 (2013), \newblock \doi{10.1103/PhysRevLett.110.126404}.

\bibitem{st2}   S. Johnston, A. Mukherjee, I. Elfimov, M. Berciu and G. A. Sawatzky, \newblock \emph{Charge Disproportionation without Charge Transfer in the Rare-Earth-Element Nickelates as a Possible Mechanism for the Metal-Insulator Transition}, \newblock Phys. Rev. Lett. 112, 106404 (2014), \newblock \doi{10.1103/PhysRevLett.112.106404}.

\bibitem{st21} B. Li, D. Louca, S. Yano, L.G. Marshall, J. Zhou, J.B. Goodenough, \newblock \emph{Insulating Pockets in Metallic LaNiO$_3$}, \newblock Adv. Electron. Mater.{\bf 2}, 150026 (2016), \newblock \doi{10.1002/aelm.201500261}.

\bibitem{st3}R.J. Green, M.W. Haverkort and G.A. Sawatzky, \newblock \emph{Bond disproportionation and dynamical charge fluctuations in the perovskite rare-earth nickelates}, \newblock Phys. Rev. {\bf B 94}, 195127 (2016), \newblock \doi{10.1103/PhysRevB.94.195127}.

\bibitem{st4}J. Shamblin et al., \newblock \emph{Experimental evidence for bipolaron condensation as a mechanism for the metal-insulator transition in rare-earth nickelates}, \newblock Nat. Commun. 9, 86 (2018),\newblock \doi{10.1038/s41467-017-02561-6}.



\bibitem{st41} M. Tyunina, M. Savinov, O. Pacherova and A. Dejneka, \newblock \emph{Small-polaron transport in perovskite nickelates}, \newblock Sci. Rep. {\bf 13}, 12493 (2023), \newblock \doi{doi.org/10.1038/s41598-023-39821}

\bibitem{st42} B. Cohen-Stead, K. Barros, R. Scalettar and S. Johnston, \newblock \emph{A hybrid Monte Carlo study of bond-stretching electron–phonon interactions and charge order in BaBiO$_3$}, \newblock npj Comput. Mater. {\bf 9}, 40 (2023), \newblock \doi{doi.org/10.1038/s41524-023-00998-6}

\bibitem{st43} M. Naamneh, E. Paris, D. McNally, Y. Tseng, W.R. Pudelko, D.J. Gawryluk,J. Shamblin, E. O’Quinn, B. Cohen-Stead, M. Shi, M. Radovic, M. Lang, T. Schmitt, S. Johnston and N. C. Plumb, \newblock \emph{Persistence of small polarons into the superconducting phase of Ba$_{1-x}$K$_x$BiO$_3$}, \newblock arXiv:2408.00401.


\bibitem{Kivelson2024} Z. Han, S. A. Kivelson, and P. A. Volkov, \newblock \emph{Quantum Bipolaron Superconductivity from Quadratic Electron-Phonon Coupling}, \newblock Phys. Rev. Lett. {\bf 132} 226001 (2024),\newblock \doi{10.1103/PhysRevLett.132.226001}.

\bibitem{Subedi2014}{
A.~Subedi, A.~Cavalleri, A.~Georges, \newblock \emph{Theory of nonlinear phononics for coherent
light control of solids}, \newblock Phys. Rev. B \textbf{89}, 220301 (2014)}.

\bibitem{Babadi2017}{
M.~Babadi, M.~Knap, I.~Martin, G.~Refael, E.~Demler, \newblock \emph{Theory of parametrically
amplified electron-phonon superconductivity}, \newblock Phys. Rev. B \textbf{96}, 014512 (2017)}.

\bibitem{John2021}{J. Sous, B. Kloss, D. M. Kennes, D. R. Reichman, and A. J. Millis, \newblock \emph{Phonon-induced disorder in dynamics of optically pumped metals from nonlinear electron-phonon coupling}, \newblock Nat. Commun. \textbf{12}, 5803 (2021).}

\bibitem{Kovac2024}{
K.~Kova\v c, D.~Gole\v z, M.~Mierzejewski, J.~Bon\v ca, \newblock \emph{Optical Manipulation of Bipolarons in a System with Nonlinear Electron-Phonon Coupling}, \newblock Phys. Rev. Lett. \textbf{132}, 106001 (2024)}.

\bibitem{Rini2007}{
M.~Rini, R.~Tobey, N.~Dean, J.~Itatani, Y.~Tomioka, Y.~Tokura, R.W.~Schoenlein, A.~Cavalleri, \newblock \emph{Control of the electronic phase of a manganite by mode-selective vibrational excitation}, \newblock Nature \textbf{449}, 72 (2007)}.

\bibitem{Hu2014}{
W.~Hu, S.~Kaiser, D.~Nicoletti, C.R.~Hunt, I.~Gierz, M.C.~Hoffmann, M.~Le Tacon, T.~Loew, B.~Keimer, A.~Cavalleri, \newblock \emph{Possible light-induced superconductivity in $\mathrm{K}_3\mathrm{C}_{60}$ at high temperature},\newblock Nat. Mater. \textbf{13}, 705 (2014)}.

\bibitem{Buzzi2020}{
M.~Buzzi, D.~Nicoletti, M.~Fechner, N.~Tancogne-Dejean,
M. A.~Sentef, A.~Georges, T.~Biesner, E.~Uykur, M.~Dressel,
A.~Henderson, T.~Siegrist, J.A.~Schlueter, K.~Miyagawa, K.~Kanoda, M.-S.~Nam, A.~Ardavan, J.~Coulthard, J.~Tindall, F.~Schlawin, D.~Jaksch, A.~Cavalleri, \newblock \emph{Photomolecular High-Temperature Superconductivity},\newblock Phys. Rev. X \textbf{10}, 031028 (2020)}.

\bibitem{2e1}J. Berakdar, 
\newblock \emph{Emission of correlated electron pairs following single-photon absorption by solids and surfaces},
\newblock Phys. Rev. {\bf B 58}, 9808 (1998),
\newblock \doi{doi.org/10.1103/PhysRevB.58.9808}

\bibitem{2e1a} R. Herrmann, S. Samarin, H. Schwabe, and J. Kirschner, 
\newblock \emph{Two Electron Photoemission in Solids},
\newblock Phys. Rev. Lett. 81, 2148 (1998),
\newblock \doi{doi.org/10.1103/PhysRevLett.81.2148}

\bibitem{2e1b}  F. O. Schumann, C. Winkler, G. Kerherve, and J. Kirschner, 
\newblock \emph{Mapping the electron correlation in two-electron photoemission},
\newblock Phys. Rev. {\bf B 73}, 041404R (2006),
\newblock \doi{doi.org/10.1103/PhysRevB.73.041404}

\bibitem{2e1c} F. O. Schumann, C. Winkler, and J. Kirschner, 
\newblock \emph{Correlation Effects in Two Electron Photoemission},
\newblock Phys. Rev. Lett 98, 257604 (2007),
\newblock \doi{doi.org/10.1103/PhysRevLett.98.257604}

\bibitem{2e2}  Yuehua Su and Chao Zhang, 
\newblock \emph{Coincidence angle-resolved photoemission spectroscopy: Proposal for detection of two-particle correlations}
\newblock Phys. Rev. {\bf B 101}, 205110 (2020),
\newblock \doi{doi.org/10.1103/PhysRevB.101.205110}

\bibitem{2e2a} T.P. Devereaux, M. Claassen, Xu-Xin Huang, M. Zaletel, J. E. Moore, D. Morr, F. Mahmood, P. Abbamonte, and Zhi-Xun Shen,
\newblock \emph{Angle-resolved pair photoemission theory for correlated electrons}
\newblock Phys. Rev. {\bf B 108}, 165134 (2023),
\newblock \doi{doi.org/10.1103/PhysRevB.108.165134}.
 
\bibitem{Schollwock2011}{U. Schollw\"{o}ck, 
\newblock \emph{The density-matrix renormalization group in the age of matrix product states}
\newblock Ann. Phys. (Amsterdam) {\bf 326}, 96 (2011).
\newblock \doi{doi.org/10.1016/j.aop.2010.09.012}}


\bibitem{Wellein1997}{G.~Wellein and H.~Fehske
\newblock \emph{Polaron band formation in the Holstein model}
\newblock Phys. Rev. B \textbf{56}, 4513 (1997).}


\bibitem{Jeckelmann1998}{E.~Jeckelmann and S.~R.~White
\newblock \emph{Density-matrix renormalization-group study of the polaron problem in the Holstein model}
\newblock Phys. Rev. B \textbf{57}, 6376 (1998).}


\bibitem{Bonca1999}{J.~Bon\v ca, S. A.~Trugman, I.~Batisti\v c, 
\newblock \emph{Holstein polaron}, 
\newblock Phys.Rev.B \textbf{60}, 1633 (1999).}


\bibitem{Barisic2002}{O.S.~Bari\v sić,
\newblock \emph{Variational study of the Holstein polaron}, 
\newblock Phys.Rev.B \textbf{65}, 144301 (2002).}


\bibitem{Hohenadler2003}{M.~Hohenadler, M.~Aichhorn, W.~von der Linden,
\newblock \emph{Spectral function of electron-phonon models by cluster perturbation theory},
\newblock Phys. Rev. B \textbf{68}, 184304 (2003).}


\bibitem{Berciu2006}{M.~Berciu,
\newblock \emph{Green’s Function of a Dressed Particle}
\newblock Phys. Rev. Lett. \textbf{97}, 036402 (2006).}


\bibitem{Mitric2022}{P.~Mitrić, V.~Janković, N.~Vukmirović, D.~Tanasković,
\newblock \emph{Spectral Functions of the Holstein Polaron: Exact and Approximate Solutions}
\newblock Phys. Rev. Lett. \textbf{129}, 096401 (2022).}


\bibitem{Mitric2023}{P.~Mitrić, V.~Janković, N.~Vukmirović, D.~Tanasković,
\newblock \emph{Cumulant expansion in the Holstein model: Spectral functions and mobility}
\newblock Phys. Rev. B \textbf{107}, 125165 (2023).}


\bibitem{Hoh}{P. C. Hohenberg,
\newblock \emph{Existence of long-range order in one
and two dimensions}
\newblock Phys. Rev.  \textbf{158}, 383 (1967).}

\bibitem{Mermin-Wagner}{N. D. Mermin and H. Wagner,
\newblock \emph{Absence of ferromagnetism or antiferromagnetism in one- or
two-dimensional isotropic heisenberg models}
\newblock Phys. Rev. Lett. \textbf{17}, 1133 (1966).}

\bibitem{Bonca2000}{J. Bon\v ca, T. Katrasnik, S.A. Trugman, 
\textit{Mobile Bipolaron}, Phy. Rev. Lett. {\bf  84}, 3153 (2000)}
  
\bibitem{Ku2002}{L. C.~Ku, S. A.~Trugman, J.~Bon\v ca, \textit{Dimensionality effects on the Holstein polaron}, Phys.Rev.B \textbf{65}, 174306 (2002).}


\bibitem{Nocera2022} A. Nocera and G. Alvarez, 
\newblock \emph{Root-N Krylov-space correction vectors for spectral functions with the density matrix renormalization group}
\newblock Phys. Rev. B {\bf 106}, 205106 (2022).
\newblock \doi{doi.org/10.1103/PhysRevB.106.205106}

\bibitem{Kuhner1999}{
T.~D. K\"{u}hner and S.~R. White,
\newblock \emph{Dynamical correlation functions using the density matrix
  renormalization group},
\newblock Phys. Rev. B \textbf{60}, 335 (1999),
		\newblock \doi{10.1103/PhysRevB.60.335}}.

\bibitem{Jeckelmann2002}{
E.~Jeckelmann,
\newblock \emph{Dynamical density-matrix renormalization-group method},
\newblock Phys. Rev. B \textbf{66}, 045114 (2002),
		\newblock \doi{10.1103/PhysRevB.66.045114}}.

\bibitem{NoceraPRE2016}{
A.~Nocera and G.~Alvarez,
\newblock \emph{Spectral functions with the density matrix renormalization
  group: Krylov-space approach for correction vectors},
\newblock Phys. Rev. E \textbf{94}, 053308 (2016),
\newblock \doi{10.1103/PhysRevE.94.053308}}.
  
\bibitem{AndreaRMP} Andrea Damascelli, Zahid Hussain, and Zhi-Xun Shen,
\newblock \emph{Angle-resolved photoemission studies of the cuprate superconductors},
\newblock Rev. Mod. Phys. {\bf 75}, 473 (2003).
\newblock \doi{doi.org/10.1103/RevModPhys.75.473}

\bibitem{SM} See Supplemental Material [url] for a description of how VED and DMRG calculations were performed. It also includes Refs. [66-81]

\bibitem{Berciu06} M. Berciu and G.L. Goodvin, \newblock \emph{Systematic improvement of the Momentum Average approximation for the Green's function of a Holstein polaron}
\newblock Phys. Rev. B {\bf 76}, 165109 (2007),
\newblock \doi{10.1103/PhysRevB.76.165109}.

\bibitem{Bonca2019}{J. Bon\v ca, S.A. Trugman, M.~Berciu, \textit{Spectral function of the Holstein polaron at finite temperature}, Phys. Rev. B {\bf   100}, 094307 (2019)}.

\bibitem{Bonca2022}J. Bon\v ca, S.A. Trugman, 
\newblock \emph{Electron removal spectral function of a polaron coupled to dispersive optical phonons}
\newblock Phys. Rev. B {\bf 106}, 174303, (2022). 
\newblock \doi{doi.org/10.1103/PhysRevB.106.174303}

\bibitem{AD2} F. Boschini, M. Zonno, and A. Damascelli,
\newblock \emph{Time-resolved ARPES studies of quantum materials}
\newblock  Rev. Mod. Phys. {\bf 96}, 015003 (2024)
\newblock \doi{doi.org/10.1103/RevModPhys.96.015003}

\bibitem{nn} N. H. Jo, E. Gati and H. Pfau
 \newblock \emph{Uniaxial stress effect on the electronic structure of quantum materials}
 \newblock 	arXiv:2405.01638 
 \newblock \doi{https://doi.org/10.48550/arXiv.2405.01638}

 
\bibitem{pp} D. Fausti , R. I. Tobey, N. Dean, S. Kaiser, A. Dienst, M. C. Hoffmann, S. Pyon, T. Takayama, H. Takagi, and A. Cavalleri , 
\newblock \emph{Light-Induced Superconductivity in a Stripe-Ordered Cuprate}
\newblock Science {\bf 331}, 6014 (2011); 
\newblock \doi{10.1126/science.1197294}

\bibitem{pp1} M. Mitrano, A. Cantaluppi, D. Nicoletti, S. Kaiser, A. Perucchi, S. Lupi, P. Di Pietro, D. Pontiroli, M. Riccò, S. R. Clark, D. Jaksch and A. Cavalleri, 
\newblock \emph{Possible light-induced superconductivity in K$_3$C$_{60}$ at high temperature}
\newblock Nature {\bf 530}, 461 (2016); 
\newblock \doi{doi.org/10.1038/nature16522}

\bibitem{pp2} S. Kaiser, C. R. Hunt, D. Nicoletti, W. Hu, I. Gierz, H. Y. Liu, M. Le Tacon, T. Loew, D. Haug, B. Keimer, and A. Cavalleri, 
\newblock \emph{Optically induced coherent transport far above T$_C$ in underdoped YBa$_2$Cu$_3$O$_{6+\delta}$}, 
\newblock Phys. Rev. {\bf B 89}, 184516 (2014). 
\newblock \doi{10.1103/PhysRevB.89.184516}.

\bibitem{pp3} G. Mazza and A. Georges, \newblock \emph{Nonequilibrium superconductivity in driven alkali-doped fullerides}, \newblock Phys. Rev. B {\bf 96}, 064515 (2017), \newblock \doi{10.1103/PhysRevB.96.064515}.

\bibitem{pp4} D.  Kennes, E. Wilner, D. Reichman and A.J. Millis. \newblock \emph{Transient superconductivity from electronic squeezing of optically pumped phonons}, \newblock. Nature Phys. {\bf 13}, 479 (2017), \newblock \doi{10.1038/nphys4024}. 


\bibitem{HHqs} F. Herrera and R. V. Krems, 
\newblock \emph{Tunable Holstein model with cold polar molecules}
\newblock Phys. Rev. A 84, 051401(R) (2011); 
\newblock \doi{doi.org/10.1103/PhysRevA.84.051401}



\bibitem{HHqsa} F. Herrera, K.W. Madison, R.V. Krems, and M. Berciu, 
\newblock \emph{Investigating Polaron Transitions with Polar Molecules}
\newblock Phys. Rev. Lett. 110, 223002 (2013);   
\newblock \doi{doi.org/10.1103/PhysRevLett.110.223002}

\bibitem{Lanczos1950}{
C. Lanczos, 
\newblock \emph{An iteration method for the solution of the eigenvalue problem of linear differential and integral operators},
\newblock J. Res. Nat. Bur. Stand. \textbf{45}, 255 (1950)}.

\bibitem{Shastry1990}{
B.S.~Shastry,  B.~Sutherland, 
\newblock \emph{Twisted boundary conditions and effective mass in Heisenberg-Ising and Hubbard rings},
\newblock Phys. Rev. Lett. \textbf{65}, 243 (1990)}.

\bibitem{Bonca2003}{
J.~Bon\v ca, P.~Prelov\v sek,
\newblock \emph{Thermodynamics of the planar Hubbard model},
\newblock Phys. Rev. B \textbf{67}, 085103 (2003)}.

\bibitem{Alvarez0209}{
G.~Alvarez,
\newblock \emph{The density matrix renormalization group for strongly
  correlated electron systems: A generic implementation},
\newblock Computer Physics Communications \textbf{180}(9), 1572 (2009)}.

\bibitem{DMRG2}{
C.~Zhang, E.~Jeckelmann and S.~R. White,
\newblock \emph{Dynamical properties of the one-dimensional holstein model},
\newblock Phys. Rev. B \textbf{60}, 14092 (1999),
\newblock \doi{10.1103/PhysRevB.60.14092}}.

\bibitem{Zhang1998}{
C.~Zhang, E.~Jeckelmann and S.~R. White,
\newblock \emph{Density matrix approach to local hilbert space reduction},
\newblock Phys. Rev. Lett. \textbf{80}, 2661 (1998),
\newblock \doi{10.1103/PhysRevLett.80.2661}}.

\bibitem{Cheng2012}{
C.~Guo, A.~Weichselbaum, J.~von Delft and M.~Vojta,
\newblock \emph{Critical and strong-coupling phases in one- and two-bath
  spin-boson models},
\newblock Phys. Rev. Lett. \textbf{108}, 160401 (2012),
\newblock \doi{10.1103/PhysRevLett.108.160401}}.

\bibitem{Brockt2015}{
C.~Brockt, F.~Dorfner, L.~Vidmar, F.~Heidrich-Meisner and E.~Jeckelmann,
\newblock \emph{Matrix-product-state method with a dynamical local basis
  optimization for bosonic systems out of equilibrium},
\newblock Phys. Rev. B \textbf{92}, 241106 (2015),
\newblock \doi{10.1103/PhysRevB.92.241106}}.

\bibitem{Jansen2021}{
D.~Jansen, C.~Jooss and F.~Heidrich-Meisner,
\newblock \emph{Charge density wave breakdown in a heterostructure with
  electron-phonon coupling},
\newblock Phys. Rev. B \textbf{104}, 195116 (2021),
\newblock \doi{10.1103/PhysRevB.104.195116}}.

\bibitem{Stolpp2021}{
J.~Stolpp, T.~K{\"o}hler, S.~R. Manmana, E.~Jeckelmann, F.~Heidrich-Meisner and
  S.~Paeckel,
\newblock \emph{Comparative study of state-of-the-art matrix-product-state
  methods for lattice models with large local hilbert spaces without u (1)
  symmetry},
\newblock Computer Physics Communications \textbf{269}, 108106 (2021)}.

\bibitem{DMRGE14}{
J.~Stolpp, J.~Herbrych, F.~Dorfner, E.~Dagotto and F.~Heidrich-Meisner,
\newblock \emph{Charge-density-wave melting in the one-dimensional {H}olstein
  model},
\newblock Phys. Rev. B \textbf{101}, 035134 (2020),
\newblock \doi{10.1103/PhysRevB.101.035134}}.

\bibitem{Jansen2022}{
D.~Jansen, J.~Bon\ifmmode~\check{c}\else \v{c}\fi{}a and F.~Heidrich-Meisner,
\newblock \emph{Finite-temperature optical conductivity with density-matrix
  renormalization group methods for the holstein polaron and bipolaron with
  dispersive phonons},
\newblock Phys. Rev. B \textbf{106}, 155129 (2022),
\newblock \doi{10.1103/PhysRevB.106.155129}}.

\bibitem{kohler2021}{
T.~K{\"o}hler, J.~Stolpp and S.~Paeckel,
\newblock \emph{Efficient and flexible approach to simulate low-dimensional
  quantum lattice models with large local hilbert spaces},
\newblock SciPost Physics \textbf{10}(3), 058 (2021)}.

\bibitem{mardazad2021}
S.~Mardazad, Y.~Xu, X.~Yang, M.~Grundner, U.~Schollw{\"o}ck, H.~Ma, and
S.~Paeckel,
\newblock \emph{Quantum dynamics simulation of intramolecular singlet fission
  in covalently linked tetracene dimer},
\newblock The Journal of Chemical Physics \textbf{155}(19), 194101 (2021).

\bibitem{Nocera2023}{
A.~Nocera, M.~Berciu,
\newblock \emph{Electron addition spectral functions of low-density polaron liquids},
\newblock SciPost Physics \textbf{15}(3), 110 (2023)}.

\bibitem{Jeckelmann2009}
T.~Shirakawa and E.~Jeckelmann,
\newblock \emph{Charge and spin Drude weight of the one-dimensional extended Hubbard model
at quarter filling},
\newblock Phys. Rev. B \textbf{79}, 195121 (2009),
\newblock \doi{10.1103/PhysRevB.79.195121}.

\end{thebibliography}

\begin{thebibliography}{99}

\bibitem{Bonča1999}{J.~Bonča, S. A.~Trugman, I.~Batistič, \textit{Holstein polaron}, Phys.Rev.B \textbf{60}, 1633 (1999).}.

\bibitem{Ku2002}{L. C.~Ku, S. A.~Trugman, J.~Bonča, \textit{Dimensionality effects on the Holstein polaron}, Phys.Rev.B \textbf{65}, 174306 (2002).}.

\bibitem{Lanczos1950}{
C. Lanczos, 
\newblock \emph{An iteration method for the solution of the eigenvalue problem of linear differential and integral operators},
\newblock J. Res. Nat. Bur. Stand. \textbf{45}, 255 (1950)}.


\bibitem{Shastry1990}{
B.S.~Shastry,  B.~Sutherland, 
\newblock \emph{Twisted boundary conditions and effective mass in Heisenberg-Ising and Hubbard rings},
\newblock Phys. Rev. Lett. \textbf{65}, 243 (1990)}.


\bibitem{Bonča2003}{
J.~Bonča, P.~Prelovšek,
\newblock \emph{Thermodynamics of the planar Hubbard model},
\newblock Phys. Rev. B \textbf{67}, 085103 (2003)}.


\bibitem{Kuhner1999}{
T.~D. K\"{u}hner and S.~R. White,
\newblock \emph{Dynamical correlation functions using the density matrix
  renormalization group},
\newblock Phys. Rev. B \textbf{60}, 335 (1999),
		\newblock \doi{10.1103/PhysRevB.60.335}}.

\bibitem{Jeckelmann2002}{
E.~Jeckelmann,
\newblock \emph{Dynamical density-matrix renormalization-group method},
\newblock Phys. Rev. B \textbf{66}, 045114 (2002),
		\newblock \doi{10.1103/PhysRevB.66.045114}}.

  
\bibitem{Nocera2022} {A. Nocera and G. Alvarez, Phys. Rev. B {\bf 106}, 205106 (2022)}.



\bibitem{NoceraPRE2016}{
A.~Nocera and G.~Alvarez,
\newblock \emph{Spectral functions with the density matrix renormalization
  group: Krylov-space approach for correction vectors},
\newblock Phys. Rev. E \textbf{94}, 053308 (2016),
\newblock \doi{10.1103/PhysRevE.94.053308}}.


\bibitem{Alvarez0209}{
G.~Alvarez,
\newblock \emph{The density matrix renormalization group for strongly
  correlated electron systems: A generic implementation},
\newblock Computer Physics Communications \textbf{180}(9), 1572 (2009)}.


\bibitem{Jeckelmann1998}{
E.~Jeckelmann and S.~R. White,
\newblock \emph{Density-matrix renormalization-group study of the polaron
  problem in the holstein model},
\newblock Phys. Rev. B \textbf{57}, 6376 (1998),
		\newblock \doi{10.1103/PhysRevB.57.6376}}.

\bibitem{DMRG2}{
C.~Zhang, E.~Jeckelmann and S.~R. White,
\newblock \emph{Dynamical properties of the one-dimensional holstein model},
\newblock Phys. Rev. B \textbf{60}, 14092 (1999),
\newblock \doi{10.1103/PhysRevB.60.14092}}.




\bibitem{Zhang1998}{
C.~Zhang, E.~Jeckelmann and S.~R. White,
\newblock \emph{Density matrix approach to local hilbert space reduction},
\newblock Phys. Rev. Lett. \textbf{80}, 2661 (1998),
\newblock \doi{10.1103/PhysRevLett.80.2661}}.

\bibitem{Cheng2012}{
C.~Guo, A.~Weichselbaum, J.~von Delft and M.~Vojta,
\newblock \emph{Critical and strong-coupling phases in one- and two-bath
  spin-boson models},
\newblock Phys. Rev. Lett. \textbf{108}, 160401 (2012),
\newblock \doi{10.1103/PhysRevLett.108.160401}}.

\bibitem{Brockt2015}{
C.~Brockt, F.~Dorfner, L.~Vidmar, F.~Heidrich-Meisner and E.~Jeckelmann,
\newblock \emph{Matrix-product-state method with a dynamical local basis
  optimization for bosonic systems out of equilibrium},
\newblock Phys. Rev. B \textbf{92}, 241106 (2015),
\newblock \doi{10.1103/PhysRevB.92.241106}}.

\bibitem{Jansen2021}{
D.~Jansen, C.~Jooss and F.~Heidrich-Meisner,
\newblock \emph{Charge density wave breakdown in a heterostructure with
  electron-phonon coupling},
\newblock Phys. Rev. B \textbf{104}, 195116 (2021),
\newblock \doi{10.1103/PhysRevB.104.195116}}.

\bibitem{Stolpp2021}{
J.~Stolpp, T.~K{\"o}hler, S.~R. Manmana, E.~Jeckelmann, F.~Heidrich-Meisner and
  S.~Paeckel,
\newblock \emph{Comparative study of state-of-the-art matrix-product-state
  methods for lattice models with large local hilbert spaces without u (1)
  symmetry},
\newblock Computer Physics Communications \textbf{269}, 108106 (2021)}.

\bibitem{DMRGE14}{
J.~Stolpp, J.~Herbrych, F.~Dorfner, E.~Dagotto and F.~Heidrich-Meisner,
\newblock \emph{Charge-density-wave melting in the one-dimensional {H}olstein
  model},
\newblock Phys. Rev. B \textbf{101}, 035134 (2020),
\newblock \doi{10.1103/PhysRevB.101.035134}}.

\bibitem{Jansen2022}{
D.~Jansen, J.~Bon\ifmmode~\check{c}\else \v{c}\fi{}a and F.~Heidrich-Meisner,
\newblock \emph{Finite-temperature optical conductivity with density-matrix
  renormalization group methods for the holstein polaron and bipolaron with
  dispersive phonons},
\newblock Phys. Rev. B \textbf{106}, 155129 (2022),
\newblock \doi{10.1103/PhysRevB.106.155129}}.


\bibitem{kohler2021}{
T.~K{\"o}hler, J.~Stolpp and S.~Paeckel,
\newblock \emph{Efficient and flexible approach to simulate low-dimensional
  quantum lattice models with large local hilbert spaces},
\newblock SciPost Physics \textbf{10}(3), 058 (2021)}.

\bibitem{mardazad2021}
S.~Mardazad, Y.~Xu, X.~Yang, M.~Grundner, U.~Schollw{\"o}ck, H.~Ma, and
S.~Paeckel,
\newblock \emph{Quantum dynamics simulation of intramolecular singlet fission
  in covalently linked tetracene dimer},
\newblock The Journal of Chemical Physics \textbf{155}(19), 194101 (2021).



\bibitem{Nocera2023}{
A.~Nocera, M.~Berciu,
\newblock \emph{Electron addition spectral functions of low-density polaron liquids},
\newblock SciPost Physics \textbf{15}(3), 110 (2023)}.

\end{thebibliography}
\end{document}